\documentclass[%
 reprint,
superscriptaddress,
 amsmath,amssymb,
 aps,
pra,
]{revtex4-2}

\usepackage{xcolor}

\usepackage{graphicx}% Include figure files
\usepackage{dcolumn}% Align table columns on decimal point
\usepackage{bm}% bold math
\usepackage{hyperref}% add hypertext capabilities
\usepackage{siunitx} % ADDED
\usepackage[utf8]{inputenc} % ADDED

\def\ZrO2{{ZrO$_2$:Y$_2$O$_3$}}

\begin{document}

\preprint{APS/123-QED}

\title{Structural color palette of disordered colloids in the Rayleigh scattering regime}% Force line breaks with \\
% \thanks{A footnote to the article title}%

\author{Kevin Vynck}
\email{kevin.vynck@univ-lyon1.fr}
\address{Institut Lumi\`{e}re Mati\`{e}re, UMR 5306, Universit\'{e} Claude Bernard Lyon 1, CNRS, 10 rue Ada Byron, 69622 Villeurbanne Cedex, France.}

\author{Amina Bensalah-Ledoux}
\address{Institut Lumi\`{e}re Mati\`{e}re, UMR 5306, Universit\'{e} Claude Bernard Lyon 1, CNRS, 10 rue Ada Byron, 69622 Villeurbanne Cedex, France.}

\author{Chirine Saadi}
\altaffiliation{Present address: Institut des Nanotechnologies de Lyon, UMR 5270, 36 avenue Guy de Collongue, 69134 Ecully Cedex, France.}
\affiliation{Institut Lumi\`{e}re Mati\`{e}re, UMR 5306, Universit\'{e} Claude Bernard Lyon 1, CNRS, 10 rue Ada Byron, 69622 Villeurbanne Cedex, France.}

\author{C\'{e}cile Le Luyer}
\affiliation{Institut Lumi\`{e}re Mati\`{e}re, UMR 5306, Universit\'{e} Claude Bernard Lyon 1, CNRS, 10 rue Ada Byron, 69622 Villeurbanne Cedex, France.}

\author{Romain Thomas}
\affiliation{Histoire des Arts et des Représentations, Université Paris Nanterre, 200 avenue de la république, 92001 Nanterre Cedex, France.}
\affiliation{Institut National d'Histoire de l'Art, 2 rue Vivienne, 75002 Paris, France.}

\author{Denis Chateau}
\affiliation{Laboratoire de Chimie, UMR 5182, ENS de Lyon, CNRS, Université Claude Bernard Lyon 1, 46 allée d’Italie, 69362 Lyon Cedex 07, France.}

\author{St\'{e}phane Parola}
\affiliation{Laboratoire de Chimie, UMR 5182, ENS de Lyon, CNRS, Université Claude Bernard Lyon 1, 46 allée d’Italie, 69362 Lyon Cedex 07, France.}

\author{Anne Pillonnet}
\email{anne.pillonnet@univ-lyon1.fr}
\affiliation{Institut Lumi\`{e}re Mati\`{e}re, UMR 5306, Universit\'{e} Claude Bernard Lyon 1, CNRS, 10 rue Ada Byron, 69622 Villeurbanne Cedex, France.}

%\date{\today}% It is always \today, today,
             %  but any date may be explicitly specified

\begin{abstract}
Structural coloration by Rayleigh scattering is widespread in nature and holds a prominent place in various art objects over a broad period of time. Beyond the common statement that Rayleigh scattering is the primary mechanism behind the multiple colored appearances of the sky, it appears that the relationship between material parameters and the colors appearing in different observation conditions has not been thoroughly explored so far. This study provides a comprehensive overview of Rayleigh scattering-based structural colors as functions of key material properties, and introduces a scalable, environmentally friendly method to fabricate solid composites with targeted colors in both reflection and transmission. Monte Carlo light transport simulations are performed to compute the structural color palette of disordered colloids -- dielectric particles in a nonscattering matrix -- in different observation modes. We provide a range of physical parameters in which the materials exhibit the same blue color in diffuse reflection and transmission. We also show that, counterintuitively, the addition of black absorbents to the matrix of a white (opaque) material can lead to the emergence of a blue coloration in diffuse reflection, thanks to the interplay between multiple scattering and absorption. Our predictions are validated by optical experiments on colloidal suspensions of Yttria-stabilized Zirconia (\ZrO2) nanoparticles in aqueous solutions. The potential of Rayleigh-scattering materials for visual arts and design is further supported by realizing solid-state composites based on abundant materials, namely borosilicate clays and hybrid silica-based glasses, using soft chemistry at room temperature.
\end{abstract}

%\keywords{Suggested keywords}%Use showkeys class option if keyword
                              %display desired
\maketitle

%\tableofcontents

\section{Introduction}

Disordered media in optics are characterized by random variations of the refractive index in space, which scramble the propagation of an incident light by scattering. Visually, a disordered medium can appear opaque, translucent or transparent, matte or glossy, and exhibit more or less pronounced colors, depending on the microstructure properties~\cite{hunter1987measurement, hebert2022optical}.

Colored disordered materials are widespread in our daily life, be they natural or manufactured, an important benefit of disorder being often the possibility to fabricate materials without expensive technological tools. Colors, as in paints or cosmetic products, are usually obtained by adding pigments or dyes to a white (nonabsorbing) material, in which case the colored appearance results from an interplay between narrowband light absorption and broadband multiple light scattering. However, colors can also be created by exploiting coherent scattering phenomena in wavelength-scale or subwavelength-scale structures, which induce a notable spectral dependence in the optical response of the material~\cite{kinoshita2008physics}. The perspective of creating dye or pigment-free materials, more easily recyclable and exhibiting colors that do not deteriorate with time, has been a great motivation for scientists in academia and industry over the past decades~\cite{goerlitzer2018bioinspired, frkapetesic2023structural}. Research on so-called structural colors nowadays largely benefits from the broad panel of available nanofabrication techniques and from the study of natural species, which serve as a source of inspiration~\cite{kinoshita2008structural, dumanli2016recent, chen2021bio}.

Diffuse (i.e., non-specular) structural colors arising from disordered nanostructures are of particular interest due to their potential in numerous fields of activity, including cosmetics, security printing, photovoltaics, radiative cooling, sensing, and decorative arts. Diffuse coloration has been achieved notably from random colloidal suspensions and amorphous packings of dielectric particles (a.k.a. photonic liquids and glasses)~\cite{takeoka2009structural, forster2010biomimetic, park2014full, schertel2019structural, okazaki2021color, demirors2024tuning, yamana2026structural}, disordered chiral nematic structures~\cite{chan2019visual, parker2022cellulose} and randomly-textured surfaces and disordered metasurfaces~\cite{song2017reproducing, vynck2022visual, agreda2023tailoring, billiet2024engineering, chen2025emergent}. In these nanostructures, color tunability is enabled by engineering the spectral resonances of individual scatterers or by implementing wavelength-scale structural correlations within the material.

Though less versatile, the most widespread and accessible mechanism for structural colors is certainly Rayleigh (or Tyndall) scattering by molecules and small particles~\cite{tyndall1869iv, rayleigh1899xxxiv, young1982rayleigh}. Spatial variations of the refractive index on scales much smaller than the wavelength $\lambda$ exhibit a scattering efficiency that varies as $\lambda^{-4}$. This makes that light in the blue part of the spectrum is more effectively scattered than light in the red part of the spectrum. In nature, the best-known example of a turbid medium in the Rayleigh scattering regime for visible light is certainly the earth atmosphere~\cite{rayleigh1899xxxiv}, exhibiting multiple colored appearances from blue to red as a result from scattering under varying lighting and observation conditions~\cite{hoeppe2007sky}. Another popular example is the iris of blue eyes~\cite{mason1924blue}, which acquires its coloration by light scattering in the iris stroma. The characteristic blue color of hydrothermal water can also be attributed to the presence of small colloidal particles in suspension~\cite{ohsawa2002rayleigh}. Let us finally point out that the literature abounds on the study of structural blue in the living world~\cite{prum2004structural, kinoshita2005structural, saranathan2012structure}, but Rayleigh scattering alone rarely explains the observed spectral behaviors -- spatial correlations in the disordered arrangement of scattering elements are often as important as the optical properties of the individual scattering elements~\cite{vynck2023light}.

In the field of materials, and in particular art objects, Rayleigh scattering was used to create blue hues long before the understanding of the underlying optical phenomenon~\cite{zhu2016effects, rio2021study, goyer2021bleus}. Blue pigments are indeed rare in nature, but obtaining a vivid structural blue requires control over the size distribution of particles and the homogeneity of the particle dispersion. In painting, for instance, the pictorial oil technique developed during the Renaissance by Flemish artists (15th century) results in a pictorial layer containing a random distribution of particles in a matrix. While colors in paintings are generally attributed to the presence of pigments, it has been demonstrated that this technique allows for blue coloration without blue pigment~\cite{goyer2019structural}. Leonardo da Vinci's manuscripts, in particular, describe the method with a judicious observation, following a discussion on the blue color of the atmosphere: ``[...] the lustre of white will nowhere display a more beautiful blue than above black - but it [the coating of white] must be very thin and finely ground''~\cite{mccurdy1938notebooks}. The size distribution of white pigment particles -- lead white -- in some studied paint layers of this time period, including da Vinci's, has been measured to be centered on 50~\unit{\nm}, corresponding well to Rayleigh scattering~\cite{gonzalez2016composition}. The intentional use of this structural staining protocol, more commonly described as the effect of turbid environments in art history, is considered by the great old masters of easel painting~\cite{gonzalez2020microchemical, goyer2021bleus}, although yellowing and cracking of varnishes over time may have altered the observation of that coloring effect today.

Structural blue coloring has also been demonstrated in ancient ceramics including the famous sky blue, sky green glazes of Chinese porcelains of the Northern Song Dynasty (960-1279), especially Jun~\cite{yang2005microstructural, zhu2016effects} and Ru~\cite{yang2005microstructural} ceramics, but also more recently in Celadon glazes~\cite{liu2022analysis}. It was shown that the heat treatment of the glaze generates a phase separation in the drops, forming a disordered network of nanoscale structures, smaller than 100~\unit{\nm}, leading to Rayleigh scattering~\cite{yang2005microstructural, zhu2016effects}. The observed green color is achieved by the addition of iron, which plays a role in the nucleation of diffusing nanophases and also acts as an optical absorber that attenuates multiple scattering in a restricted spectral region.

The application of scientific methodologies enabled the production of various materials colored by Rayleigh scattering. Let us mention quasi-amorphous particle films~\cite{ge2014angle, xiang2016new}, ceramic glasses~\cite{rio2021study} and aerogels~\cite{Longo2014aerogel, michaloudis2024sky}. In these studies, the colors have been obtained empirically. As might be unexpected, to our knowledge, the literature still lacks a comprehensive and systematic study of the structural colors achievable by Rayleigh scattering, thereby hindering the practical design of materials with targeted colored appearances.

\begin{figure*}[ht!]
\centering
  \includegraphics[width=151.293mm]{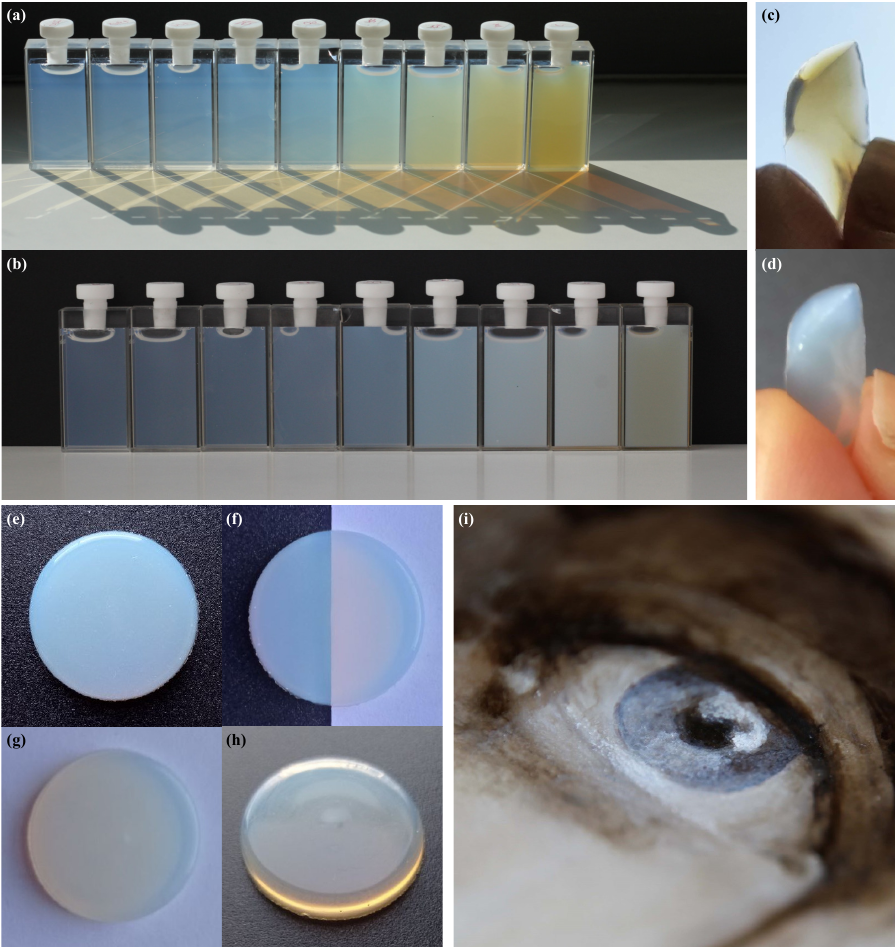}
  \caption{Panoply of liquid and solid Rayleigh-scattering materials, observed in different realistic conditions, often yielding different perceived colors. (a)-(b) Photographs of aqueous solutions containing 20-nm-diameter \ZrO2 particles with increasing particle concentrations (from left to right). (c)-(d) Photographs of borosilicate clay composite containing 20-nm-diameter \ZrO2 particles. (e)-(h) Photographs of a hybrid silica-based glass composite containing 20-nm-diameter \ZrO2 particles. (i) Photograph of a pictorial film producing a blue color without blue pigment [``Hineni'' (2018) by Anne Goyer, exposed at the Pio Monte della Misericordia's museum in Naples, Italy; adapted with permission from Ref.~\cite{goyer2021bleus}]. Details on the dimensions and composition of the Rayleigh-scattering materials are provided in the text.}
  \label{fig:solid-composite-materials}
\end{figure*}

In this article, we theoretically explore the palette of structural colors accessible in disordered colloidal materials in the Rayleigh scattering regime and experimentally demonstrate the validity of our predictions by fabricating liquid and solid materials exhibiting the desired colored appearances. Figure~\ref{fig:solid-composite-materials} presents a panoply of such liquid and solid Rayleigh-scattering materials in different observation conditions.

The scattering media under investigation are composed of spherical nonabsorbing dielectric particles randomly dispersed in a transparent (nonscattering) matrix. The framework of Rayleigh scattering allows us to establish general behaviors that hold for all colloidal suspensions of small particles made of non- or weakly-dispersive materials. Using a Monte Carlo method to simulate light transport in disordered media, we predict the structural colors observed in reflectance and transmittance through slabs of finite thickness as a function of the material parameters. The relevance of this palette is demonstrated by the design and optical analysis of the corresponding materials prepared by soft chemistry in aqueous solutions with high refractive index \ZrO2 particles [Fig.~\ref{fig:solid-composite-materials}(a)-(b)]. The achievement of these colors in solid matrices is demonstrated using low energy cost methods based on glasses by the sol-gel route and borosilicate clays realized at room temperature [Fig.~\ref{fig:solid-composite-materials}(c)-(h)]. The effect of adding black absorbents - carbon graphite particles in the present study - to the matrix on the resulting structural colors is finally investigated numerically and experimentally.

The potential impact of this study is two-fold: First, it may help the prediction and design of pigment-free colored turbid media to realize artworks [e.g., Fig.~\ref{fig:solid-composite-materials}(i)] and architectural creations such as smart windows that give rise to different luminous atmospheres depending on lighting conditions. Millimeter or centimeter-thick glass composites can be elaborated to offer the desired diffuse colors in both reflection and transmission while providing mechanical rigidity. In addition, the color palette of Rayleigh-scattering materials is essentially unaffected by particle size polydispersity, thereby making this solution easily scalable. Second, this study may facilitate the analysis, preservation and restoration of ancient art objects containing Rayleigh-scattering materials, thereby making a meaningful contribution to art history and cultural heritage.

\section{Structural color palette in the Rayleigh scattering regime}

The Monte Carlo method for light transport~\cite{veach1998robust} is a practical and insightful numerical approach to predict the colored appearance of macroscopic colloidal media. This method, which relies on the theoretical grounds of radiative transfer~\cite{chandrasekhar1960radiative} and plays a major role in atmospheric optics~\cite{thomas2002radiative}, biomedical optics~\cite{wang2007biomedical} and computer graphics~\cite{pharr2023physically} for instance, allows the computation of ensemble-averaged quantities (e.g., the reflectance and transmittance of a material slab, the average electromagnetic energy density within it, etc.), by random-walk simulations. Importantly, in statistically homogeneous and dilute suspensions of colloidal particles, as considered in the present study, the random-walk parameters can be easily determined from the physical parameters of the turbid medium~\cite{carminati2021principles}, thereby establishing a direct relationship between the microstructure of the medium and its optical properties on a macroscopic scale. The Monte Carlo method was used previously to explore the diffuse colors of disordered packings of Mie-resonant spherical particles in reflection~\cite{hwang2021designing}. A similar approach, detailed in the Methods section, is applied here to predict the structural colors of Rayleigh-scattering particles in reflection and transmission.

\begin{figure}[ht!]
\centering
  \includegraphics[width=79.736mm]{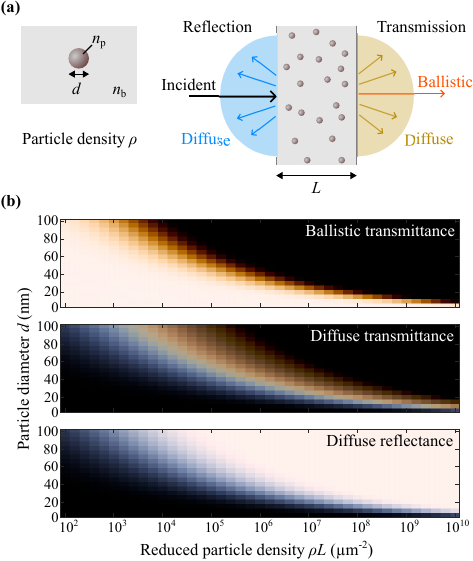} % max width = 83 mm
  \caption{Structural color palettes of nonabsorbing turbid media in the Rayleigh scattering regime. (a) Problem under consideration and definitions. Small spherical particles of diameter $d$ and refractive index $n_\text{p}$ are randomly dispersed in matrix of refractive index $n_\text{b}$ at a number density $\rho$. The turbid medium is a laterally-infinite slab of thickness $L$ and is illuminated at normal incidence. We compute and measure the ballistic transmittance, diffuse transmittance and diffuse reflectance over the visible spectrum. The diffuse components are integrated over the hemisphere. (b) Predicted structural colors for \ZrO2 particles in water, decomposed into ballistic transmittance (top), diffuse transmittance (middle) and diffuse reflectance (bottom), as a function of the reduced particle density $\rho L$ and particle diameter $d$.}
  \label{fig:numerical-palette-Rayleigh}
\end{figure}

To illustrate the range of structural colors achievable in the Rayleigh scattering regime, we simulate the optical response of colloidal suspensions of spherical, nonabsorbing \ZrO2 nanoparticles in water. The particles are distributed randomly at a number density $\rho$ in a slab of finite thickness $L$ (similar to a cuvette) and infinite lateral size placed in air. Simulations are performed assuming unpolarized light at normal incidence on the sample surface and the optical response is decomposed into three distinct observation modes: the ballistic transmittance (light being transmitted through the medium without being scattered) and the angularly-integrated diffuse transmittance and reflectance (light being transmitted or reflected after at least 1 scattering event), see Fig.~\ref{fig:numerical-palette-Rayleigh}(a).

The structural color palettes observed in the different observation modes are given in Fig.~\ref{fig:numerical-palette-Rayleigh}(b) as a function of the particle diameter $d$ and the reduced particle number density $\rho L$. The parameter $\rho L$ can be interpreted as the number of particles counted in a unit surface area when sweeping through a material of thickness $L$. In the framework of the present model, the optical response (at a given wavelength) does not vary when $\rho L$ is fixed.

The ballistic transmittance [upper panel in Fig.~\ref{fig:numerical-palette-Rayleigh}(b)] exhibits a variation between white (high transparency) and black (high extinction) with a transient region with orange/red hues, similar to a sunset. Blue colors cannot be observed, as expected, because light at shorter wavelengths is more efficiently scattered out of the ballistic direction than light at longer wavelengths.

More interestingly, the diffuse transmittance [middle panel in Fig.~\ref{fig:numerical-palette-Rayleigh}(b)] displays an atypical color variation between blue and red hues surrounded by two regions of very weak diffuse transmitted light (black). For the smaller particle sizes and lower densities, light is very weakly scattered and remains mostly in the ballistic component. Increasing these parameters, light at the shorter wavelengths experiences one or a few scattering events and essentially none at the longer wavelengths, leading to a relatively high diffuse transmission with a blue coloration. Upon further increase, one enters the multiple scattering regime where the medium becomes opaque (i.e., strongly reflective) for light at the shorter wavelengths but not yet for light at the larger wavelengths, resulting in diffuse red hues in transmission. Finally, for the higher densities and larger particle, the medium is opaque at all wavelengths (for both the ballistic and diffuse components).

The diffuse reflectance [bottom panel in Fig.~\ref{fig:numerical-palette-Rayleigh}(b)] exhibits an opposite behavior compared to the ballistic transmittance. As expected, very little light is scattered in reflection for the smaller particle sizes and lower densities, whereas in the other limit, the medium becomes opaque at all wavelengths in the visible range, resulting in a white appearance. The typical Rayleigh blue coloration appears again in an intermediate regime.

All in all, these first simulations show that the typical Rayleigh blue coloration can be observed for sets of microstructural parameters (particle size $d$, particle density $\rho$ and sample thickness $L$) but these parameters need to be finely controlled. Remarkably, in a rather narrow range of parameters, the blue coloration may be observed simultaneously in both the diffuse transmittance and reflectance, along with a high ballistic transmittance with weak orange/red hues. In terms of visual appearance, this will result in a translucent material with a blue, diffuse coloration.

\begin{figure*}[ht!]
\centering
  \includegraphics[width=150.193mm]{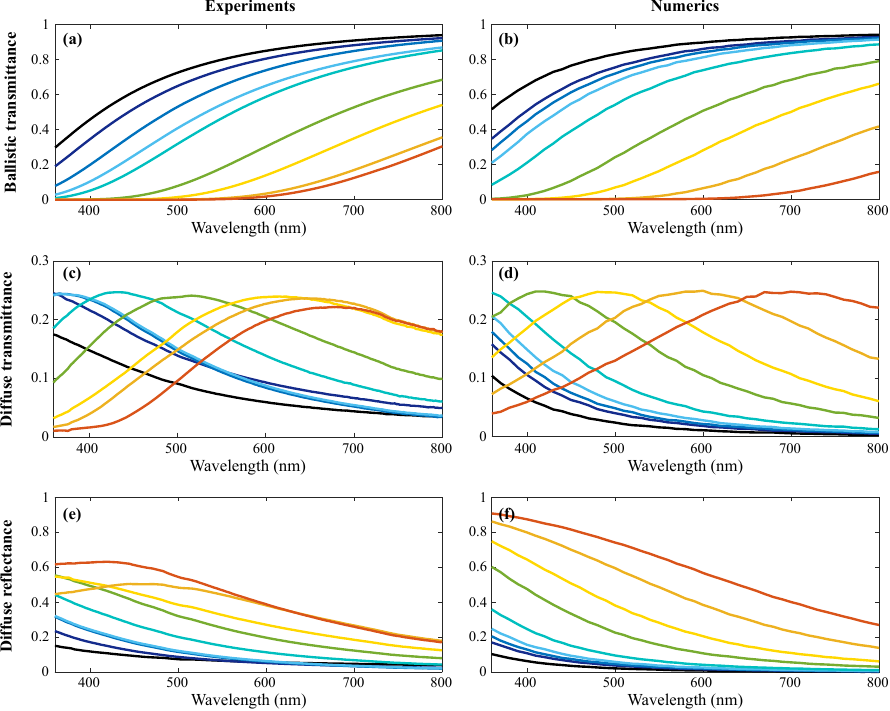} % Max width = 171 mm
  \caption{Comparison between experimental spectra (a,c,e) and numerical spectra (b,d,f) on the ballistic transmittance (a-b), diffuse transmittance (c-d) and diffuse reflectance (e-f) for samples with different particle concentrations and a sample thickness $L=10~\unit{mm}$. The solid curves with colors ranging from black to red correspond, respectively, to colloidal solutions with \ZrO2 mass fractions $[0.43, 0.71, 0.85, 1.1, 1.7, 4.1, 8.0, 18, 50]~\unit{wt\%}$ in the experiments and particle number densities $\rho = [1.71 \; 10^2, 2.84 \; 10^2, 3.41 \; 10^2, 4.26 \; 10^2, 6.82 \; 10^2, 1.71 \; 10^3, 3.41 \; 10^3, 8.53 \; 10^3, 3.41 \; 10^4]~\unit{\um^{-3}}$ in the numerics. The Monte Carlo simulations were performed here with no adjustable parameter, considering the finite lateral size of the cuvettes ($20~\unit{mm} \times 30~\unit{mm}$) and short-range structural correlations in the particle positions (for the highest \ZrO2 mass fraction of 50 wt\%, the volume filling fraction is about 14.3\%).}
  \label{fig:spectra-exp-vs-num}
\end{figure*}

\begin{figure}[ht!]
\centering
  \includegraphics[width=77.049mm]{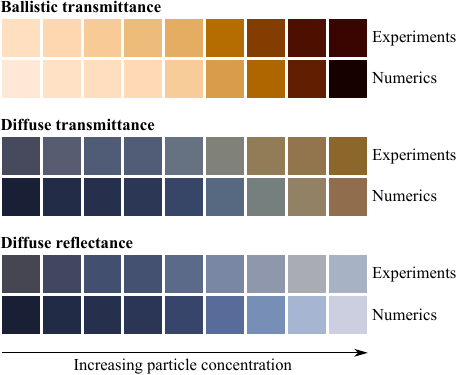} % Max width = 83 mm
  \caption{Comparison of the structural colors deduced from the experimentally measured and numerically computed spectra reported in Fig.~\ref{fig:spectra-exp-vs-num} on the wavelength range $[0.36,0.80]~\unit{\um}$. The same color variations are observed on all observation modes.}
  \label{fig:colors-exp-vs-num}
\end{figure}

We proceed by demonstrating experimentally the predicted color palettes. Towards this aim, we prepared aqueous solutions of commercial \ZrO2 particles with an average size $d = 20 \pm 3~\unit{nm}$, as estimated by Transmission Electron Microscopy (TEM), see the Methods section. Water was chosen as the matrix because it allows excellent variability in particle concentration (by adjusting the dilution) and medium thickness (by adjusting the container size). The percentage by weight of \ZrO2 particles in water in the starting solution is $50~\unit{wt\%}$, equivalent to about $3.41 \; 10^4$ particles per $\unit{\um^3}$ for a mass density of $6~\unit{g.cm^{-3}}$. We recorded the ballistic transmittance, diffuse transmittance and diffuse reflectance spectra as a function of the particle concentration for normally-incident light on cuvettes of thickness $L=10~\unit{mm}$. Monte-Carlo simulations were performed using these nominal parameters, considering also the finite lateral size of the cuvettes, which induces some leakage from the sides, as well as short-range structural correlations in the particle positions, which affect the light transport parameters for the samples with the higher particle densities, see the Methods section. The experimental and numerical spectra are presented in Fig.~\ref{fig:spectra-exp-vs-num}. Figure~\ref{fig:colors-exp-vs-num} shows the corresponding colors, obtained by converting the spectra to the sRGB color space assuming a D65 illuminant.

A good qualitative agreement is found overall between the experimental measurements and the numerical predictions; the same trends are observed for all observations modes. A slight shift of the color palette between experiments and simulations is observed. As shown in the Supporting Information, this may be due to the particle size polydispersity, disregarded in the simulations of Figs.~\ref{fig:spectra-exp-vs-num}-\ref{fig:colors-exp-vs-num}, which tends to increase the scattering efficiency of the particle dispersion without affecting the spectral dependence. Apart from this, discrepancies appear for the samples at the higher particle densities, especially in diffuse reflectance. This may be explained first by technical limitations of the setup to collect the entirety of the diffuse light, and second by the fact that the real and the modeled colloidal suspensions differ. Indeed, while modeling considers a simple fluid of identical hard spheres at equilibrium, the particles in the higher-density colloidal suspensions may clump together, thereby altering significantly their scattering properties. Nevertheless, the experiments demonstrate the possibility to achieve a blue diffuse color in reflection and transmission on a specific range of parameters from design.

\begin{figure}[ht!]
\centering
  \includegraphics[width=80.667mm]{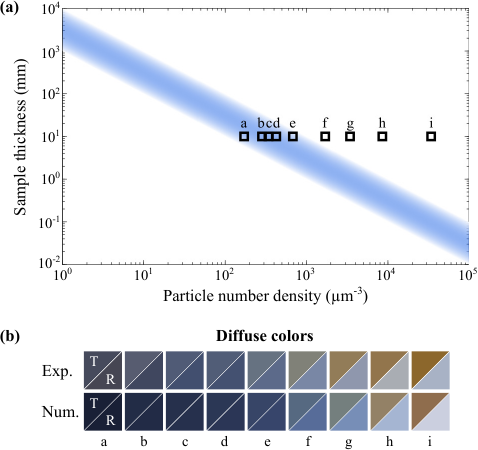}
  \caption{Isotropic blue diffuse color with disordered colloids. (a) Range of sample thicknesses for $20~\unit{nm}$-diameter \ZrO2 particles in water to obtain a blue diffuse color in both transmission and reflection (shaded area) as a function of the particle number density $\rho$. The squares, labeled from a to i, indicate the nominal parameters of the samples studied in Fig.~\ref{fig:spectra-exp-vs-num}. (b) Structural colors (already reported in Fig.~\ref{fig:colors-exp-vs-num}) for the diffuse transmittance (top-left in each square) and diffuse reflectance (bottom-right in each square). The isotropic blue diffuse color is observed in the expected range of parameters.}
  \label{fig:physical-parameters}
\end{figure}

Figure~\ref{fig:physical-parameters} offers a different viewpoint on this aspect. In Fig.~\ref{fig:physical-parameters}(a), we highlight the range of material thicknesses $L$ and particle concentrations $\rho$ for particle diameter of $20~\unit{nm}$ that yields a significant blue diffuse color simultaneously in reflection and transmission, as predicted by the numerical simulations. The more the particle concentration increases, the smaller should the sample thickness be. This reflects the fact that the relevant parameter is $\rho L$, as discussed previously. This also implies that a blue diffuse color can always be achieved at low particle densities simply by using thicker materials. For colloidal dispersions of $20~\unit{nm}$-diameter \ZrO2 particles in water, $\rho L$ should be comprised approximately between $10^6$ and $10^7 ~\unit{\um^{-2}}$, corresponding to thicknesses between $30$ and $300~\unit{\um}$ for samples at $50~\unit{wt\%}$ (i.e., $\rho=3.41 \; 10^4~\unit{\um^{-3}}$), and between $5$ and $50~\unit{mm}$ for samples at $0.5~\unit{wt\%}$ (i.e., $\rho=2 \; 10^2~\unit{\um^{-3}}$). The diffuse colors in transmission and reflection, derived from the experimental measurements for $10~\unit{mm}$-thick samples at different concentrations, are shown in Fig.~\ref{fig:physical-parameters}(b) [top-left and bottom-right parts of each square for the diffuse transmittance and reflectance, respectively]. The blue coloration readily appears when entering the desired parameter range predicted by the simulations.

\section{Solid composite material colored by Rayleigh scattering}

To demonstrate the feasibility of making solid materials colored by Rayleigh scattering without dyes or pigments, as well as the environmental relevance of this coloring process, we realized composite materials based on borosilicate clays and hybrid silica-based glasses by soft chemistry at room temperature. The 20-\unit{nm}-diameter \ZrO2 particles were incorporated during the fabrication process. The materials can easily be molded in liquid phase as macroscopic objects of various shapes and sizes. We refer to the Methods section for details on fabrication.

The clay composite was made by dissolving borosilicate powder in water. Upon drying at room temperature, we obtained solid-state samples of millimeter thicknesses with $\rho L \approx 10^{8}~\unit{\um^{-2}}$. A sample shown in Fig.~\ref{fig:solid-composite-materials}(c)-(d) exhibits an orange diffuse color in transmission and a blue diffuse color in reflection, as expected from our preditions in Fig.~\ref{fig:numerical-palette-Rayleigh}(b). An important drawback of clay composites for applications is, however, their friability.

The hybrid silica-based glass composite, much harder and easier to handle, was made by a sol-gel process at room temperature, followed by slow drying at 45$^\circ$C. We realized $4 \; \unit{mm}$-thick disk-shaped samples with a nominal particle density $\rho = 4 \; 10^3~\unit{\um^{-3}}$, corresponding to $\rho L = 1.2 \; 10^7~\unit{\um^{-2}}$. Figure~\ref{fig:solid-composite-materials}(e)-(h) shows several photographs of the same sample in different settings. On a black diffuse substrate [Fig.~\ref{fig:solid-composite-materials}(e)], the sample exhibits a characteristic blue diffuse color in reflection. When placed on a brighter reflective substrate [Fig.~\ref{fig:solid-composite-materials}(g)], the sample acquires orange/red hues in reflection. This phenomenon can be understood by considering that, in these conditions, a significant fraction of the light observed in reflection has been ballistically transmitted through the sample, reflected by the substrate and transmitted again through the sample. The resulting visual effect would not exist if blue pigments or dyes were used, precisely because light at the longer wavelengths (red part of the spectrum) would be absorbed. This is one manifestation of the atypical visual appearance of Rayleigh-scattering materials, or structurally-colored materials in general, compared to absorption-based colored disordered materials.

To conclude this section, let us recall that the Rayleigh blue color can also be achieved by a hand-crafted painting protocol, similar to that of glazing, without calibrated products or technological tools~\cite{goyer2019structural}. An illustration is given in Fig.~\ref{fig:solid-composite-materials}(i), where the Rayleigh scattering material is a paint layer composed of white lithopone grains, whose refractive index is close to zirconia, randomly dispersed in a transparent oil vanish matrix, applied on a dark bitumen coating. The turbid layer has a thickness of a few tens of micrometers and the scattering elements have a mean particle diameter of about 100~\unit{nm}. This blue color had been obtained empirically through a research-creation project~\cite{goyer2019structural}. The present study now provides some guidelines to achieve desired blue hues using Rayleigh scattering.

\section{Effect of adding black absorbents to the matrix}

A structurally-colored material naturally becomes whiter in reflection as its thickness is increased [see, e.g., the bottom panel of Fig.~\ref{fig:numerical-palette-Rayleigh}(b) at increasing values of $\rho L$ in the specific case of Rayleigh-scattering materials]. Mitigating structural color bleaching in reflection has a practical interest in visual arts (e.g., painting~\cite{goyer2019structural}), where control over the material thickness is often limited. A well-accepted strategy to reaching this aim is to add black absorbing species to the material, see for instance Refs.~\cite{forster2010biomimetic, xiang2016new, takeoka2018structural, nakamae2021thermally}. Effectively, similar to using a thinner material, this tends to suppress the longer multiple-scattering trajectories in the material, thereby enhancing the importance of the shorter trajectories that contribute to producing structural color.

In this section, we systematically analyze the impact of adding black absorbing elements to a Rayleigh-scattering material on the resulting structural colors in reflection and transmission. Monte Carlo simulations are performed for ideal (non-dispersive and nonabsorbing) Rayleigh scatterers incorporated in an absorbing matrix (see details in the Methods section). The two relevant quantities here are the scattering mean free path $\ell_\text{s}$, which describes the average distance between two scattering events in the nonabsorbing material, and the absorption mean free path $\ell_\text{a}$, which describes the average distance after which light is absorbed. For the sake of understanding, we assume that the absorption mean free path is constant throughout the entire spectrum, keeping in mind that real pigments and dyes always exhibit some spectral dependence.

\begin{figure}[tp!]
\centering
  \includegraphics[width=76.114mm]{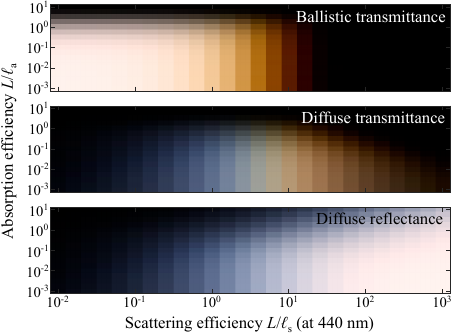}
  \caption{Structural color palette of turbid media in the Rayleigh scattering regime including black absorbing species. Similarly to Fig.~\ref{fig:numerical-palette-Rayleigh}, the optical response is decomposed into ballistic transmittance (top), diffuse transmittance (middle) and diffuse reflectance (bottom). For the sake of generality, the predicted colors are displayed as a function of the scattering efficiency $L/\ell_\text{s}$, where the scattering mean free path $\ell_\text{s}$ is taken at a wavelength of $440~\unit{nm}$, and absorption efficiency $L/\ell_\text{a}$, where the absorption mean free path $\ell_\text{a}$ is assumed to be constant over the visible spectrum. A refractive index contrast of 1.33 between the material and the ambient medium (e.g., water and air) is assumed for reflection and refraction at the material interfaces.}
  \label{fig:color-palette-absorption}
\end{figure}

The computed color palettes in the three observation modes (ballistic transmittance, diffuse transmittance and diffuse reflectance) are displayed in Fig.~\ref{fig:color-palette-absorption} as a function of the scattering efficiency $L/\ell_\text{s}$, evaluated at a wavelength $\lambda=440~\unit{nm}$, and the absorption efficiency $L/\ell_\text{a}$ in log-log scale. Let us first emphasize that, for a dilute dispersion of ideal Rayleigh scatterers, $\ell_\text{s}$ always varies as $\lambda^4$ (up to a prefactor that depends on the microstructural parameters) and single scattering is isotropic. This implies that, in absence of absorption, the dimensionless quantity $L/\ell_\text{s}$ fully determines the reflection and transmission properties of a material. This explains why the color variations in Fig.~\ref{fig:numerical-palette-Rayleigh}(b) along one axis or the other are always the same. The nonabsorbing case of Fig.~\ref{fig:numerical-palette-Rayleigh}(b) can in fact be reduced to a single horizontal line of the palette in Fig.~\ref{fig:color-palette-absorption}, in the limit where $L/\ell_\text{a}\rightarrow 0$.

A first observation on Fig.~\ref{fig:color-palette-absorption} is that the blue coloration in diffuse transmittance and reflectance [middle and bottom panels in Fig.~\ref{fig:color-palette-absorption}] appears for $L/\ell_\text{s} \sim 1$ (at $\lambda = 440~\unit{nm}$), that is, when light at the shorter wavelengths in the visible spectrum is scattered once (or at most a few times) on average, confirming our previous assertion.

Second, on the effect of absorption in the matrix, we observe an expected trend for the ballistic and diffuse transmittances [top and middle panels in Fig.~\ref{fig:color-palette-absorption}] with a darkening of the observed colors with increasing absorption efficiency. Fewer light at all wavelengths reaches the material surface in transmission.

Third and more interestingly, a blue coloration appears in the diffuse reflectance in the multiple scattering regime ($L/\ell_\text{s} \gg 1$) and for a sufficiently high absorption efficiency ($L/\ell_\text{a}>1$). In this situation, where light at all wavelengths would be essentially reflected by multiple scattering, absorption is more effective on the longer trajectories, that is, at longer wavelengths, leaving only the light at the shorter wavelengths to escape from the material before being absorbed. Thus, absorption in the matrix of disordered colloids of Rayleigh scatterers can lead to a blue coloration even in the deep multiple scattering regime. Visually, the materials would, in this case, appear very opaque (very weak transmission) yet with a blue diffuse color in reflection.

To validate this observation experimentally, we now consider a colloidal solution of $90~\unit{nm}$-diameter \ZrO2 particles dispersed in water at a volume particle density $\rho=4 \; 10^2~\unit{\um^{-3}}$ and examine how the diffuse reflectance spectrum is modified upon the addition of a black absorbent to the solution. Specifically, we use a carbon graphite particles, which are widely used in the field of art and design. This choice is further motivated by our ambition to move towards ecological materials. As shown in the Methods section, the particles do not lead to a uniform absorption efficiency in the visible range and produce additional light scattering related to their micrometer size. Therefore, our comparison between experimental and numerical results can only be qualitative.

\begin{figure}[tp!]
\centering
  \includegraphics[width=78.171mm]{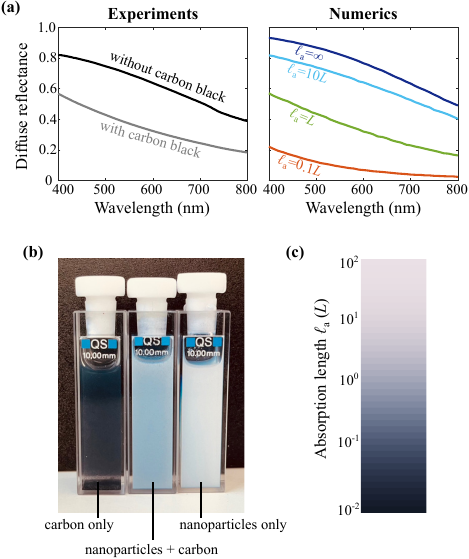}
  \caption{Qualitative comparison between the diffuse reflectances and the associated colors of disordered colloids with different degrees of absorption in the matrix. The materials consist of 90~\unit{nm}-diameter \ZrO2 particles dispersed in water, filling a cuvette of thickness $10~\unit{mm}$ and lateral side length $20~\unit{mm}$. (a) Experimental spectra of the diffuse reflectances for the samples without (black curve) and with (gray curve) carbon graphite particles (left) and numerical spectra of the same observable for materials with $L/\ell_\text{s}=50$ at $\lambda=440~\unit{nm}$ and varying absorption mean free paths $\ell_\text{a}=[\infty, 10L, L, 0.1L]$, i.e., $L/\ell_\text{a}=[0, 0.1, 1, 10]$. The numerical simulations take into account the finite lateral size of the cuvettes. (b) Photographs of the cuvettes filled with solutions containing either carbon graphite particles only (left), \ZrO2 particles only (right), or both (middle). The blue coloration, which stems from an interplay between multiple light scattering and absorption, is evident. (c) Numerical predictions of the angle-integrated color in diffuse reflectance of materials with $L/\ell_\text{s}=50$ at $\lambda=440~\unit{nm}$, as a function of the absorption length $\ell_\text{a}$ in units of the sample thickness $L$, revealing the various shades of blue observed at the transition between a fully scattering (white) sample and a fully absorbing (black) sample.}
  \label{fig:carbon-black-solutions}
\end{figure}

The left panel of Fig.~\ref{fig:carbon-black-solutions}(a) shows the measured diffuse reflectance of the colloidal solutions containing 90~\unit{nm}-diameter \ZrO2 particles at a density $\rho = 4 \; 10^2~\unit{\um^{-3}}$ and the same colloidal solution doped with carbon graphite particles at $0.3 \; 10^{-3}~\unit{g.cm^{-3}}$. The reflectance naturally diminishes at all wavelengths in the visible range but, more importantly, the spectral variation is altered, giving more weight to shorter wavelengths over longer wavelengths. The right panel of Fig.~\ref{fig:carbon-black-solutions}(a) shows the spectra computed for a material with $L/\ell_\text{s}=50$ at $\lambda = 440~\unit{nm}$ for different values of $L/\ell_\text{a}$. The spectral dependence observed experimentally is nicely reproduced. The effect of adding a black absorbent to the colloidal solution on the visual appearance of the materials is shown in Fig.~\ref{fig:carbon-black-solutions}(b). While the samples containing the carbon graphite particles only and the \ZrO2 particles only appear black and white, respectively, a blue coloration emerges when both particle types are mixed. This effect, which may be counterintuitive at first, is fully explained by the analysis above. For completeness, Fig.~\ref{fig:carbon-black-solutions}(c) finally shows the different shades of blue that can be observed in reflection when adding black absorbents to a optically-thick, white Rayleigh-scattering material.

\section*{Conclusion}

Structural coloration by Rayleigh scattering is ubiquitous in nature and in visual arts. Yet, it appears that the literature still lacked a thorough investigation of how the microstructure parameters affect the perceived colors.

One objective of the present study was to partly fill this gap by unveiling, thanks to light scattering theory and light transport simulations, the range of structural colors achievable in reflection and transmission with colloidal suspensions of small dielectric particles. Microstructural parameters, namely particle size, density, sample thickness, as well as absorption induced by black absorbing species in the matrix, were varied over several orders of magnitude to identify parameter ranges that would lead to targeted colored appearances. These include an identical blue diffuse color appearing simultaneously in reflection and transmission, or a blue diffuse color emerging in reflection from a strongly-scattering material, thanks to an interplay between (multiple) Rayleigh scattering and broadband light absorption. In spite of being derived specifically for dilute, uncorrelated assemblies of tiny dielectric spherical particles, the color palette displayed in Fig.~\ref{fig:color-palette-absorption} puts forward the primary role of two dimensionless quantities, $L/\ell_\text{s}$ and $L/\ell_\text{a}$ (where $L$, $\ell_\text{s}$ and $\ell_\text{a}$ are the sample thickness, the scattering mean free path and the absorption mean free path, respectively) on the structural colors. Supposedly, this theoretical structural color palette should give general trends for virtually any Rayleigh-scattering material containing black absorbents.

A second objective of the present study was to demonstrate our capability to realize liquid and solid materials exhibiting a targeted colored appearance [Fig.~\ref{fig:solid-composite-materials}]. We realized a series of colloidal suspensions of \ZrO2 particles in water at different particle densities, with or without the addition of carbon graphite particles in the solution, and found a good agreement with our numerical predictions on the colors they produced. An important outcome of our work is yet the fabrication of solid Rayleigh-scattering materials based on borosilicate clays and hydrid silicate glasses using soft chemistry. The objects can be realized at room temperature using abundant materials, and their shape and size can be controlled easily during preparation. Glass composites of millimeter or centimeter-scale thicknesses are mechanically rigid, which is a prerequisite for many applications. All in all, by combining theoretical predictions of structural colors and soft chemistry techniques for material fabrication, we believe that our study opens interesting perspectives for the conception and eco-friendly realization of Rayleigh-scattering materials for many branches of visual arts, including painting, decorative arts and architecture.

\begin{figure}[tp!]
\centering
  \includegraphics[width=76.911mm]{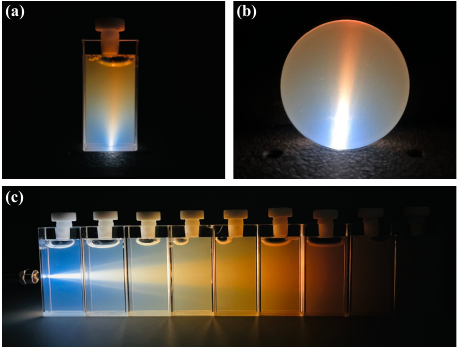}
  \caption{Multicolored appearance of Rayleigh-scattering materials containing 20-\unit{nm}-diameter \ZrO2 particles under artificial, white-light illumination. (a) Aqueous solution with a fixed particle density, (b) solid hybrid silica-based glass with a fixed particle density, and (c) series of aqueous solutions with increasing particle density (from left to right).}
  \label{fig:RayleighShadesNight}
\end{figure}

Design-wise, Rayleigh-scattering materials can not only reproduce the pleasant color variations of the sky [Fig.~\ref{fig:solid-composite-materials}(a)], but also create stunning effects, such as those displayed in Fig.~\ref{fig:RayleighShadesNight}, where liquid and solid samples are illuminated from the side by artificial white light. Specifically, the interplay between Rayleigh scattering and leakage from the sample surface results into a multicolored appearance, which could be implemented in glass-made architectural structures, for instance.

Unlike colloidal suspensions and amorphous packings of dielectric particles in the Mie-scattering regime~\cite{takeoka2009structural, forster2010biomimetic, park2014full, schertel2019structural, okazaki2021color, demirors2024tuning, yamana2026structural}, Rayleigh-scattering materials are robust to particle polydispersity. Indeed, structural colors based on Mie scattering stem from the presence of spectral resonances in the visible range and are therefore intrinsically sensitive to polydispersity. Instead, because all Rayleigh scatterers exhibit the same spectral dependence for their scattering efficiency -- the particle size only affects the prefactor, to the leading order -- the structural color palette of Rayleigh-scattering materials is essentially unaffected by polydispersity. This is shown in the Supporting Information, where we compare dilute colloidal suspensions of \ZrO2 particles with an average diameter of 20~\unit{nm} and Silicon particles with an average diameter of 160~\unit{nm}, both at the same degrees of polydispersity. The robustness of Rayleigh-scattering materials against polydispersity relaxes stringent particle sizing requirements, making them particularly amenable to low-tech manufacturing.

As a final note, let us emphasize that our results are naturally at the disposal of scientific communities that analyze Rayleigh-scattering materials, be they natural (atmospheres, glaciers, living organisms) or artificial (art objects). Indeed, the structural color palette can be used to predict the initial color of a given turbid medium, such as a pictorial layer, before its chemical or mechanical alteration. This constitutes a promising approach for the preservation and restoration of artworks, as well as for research on historical know-how in cultural heritage.

\section*{Methods}

\subsection*{Numerical simulations}

We considered a disordered assembly of identical, impenetrable spherical particles of diameter $d$ and refractive index $n_\text{p}$ in a homogeneous background with refractive index $n_\text{b}$ at a volume particle density $\rho$. The general methodology to compute the structural colors of disordered colloids was as follows. First, we computed the wavelength-dependent light transport parameters of disordered assemblies of particles from the knowledge of the particle's size, material, density and spatial organization (structural correlation). Then, we performed Monte Carlo light transport simulations in finite-size materials using those parameters to obtain the reflectance and transmittance spectra. Finally, these spectra were converted into a color assuming a standard illuminant.

\emph{Determination of light transport parameters:} The key parameters for light transport in turbid media are the scattering mean free path $\ell_\text{s}$, describing the average distance between two scattering events, and the scattering anisotropy parameter $g$, describing the angular anisotropy of the scattering diagram upon a single scattering event. Electromagnetic scattering theory establishes a direct relationship between the microstructure parameters (here, the particle size, material, density and spatial organization) and light transport parameters via the introduction of two important quantities: the differential scattering cross-section $\text{d}\sigma_\text{s}/\text{d}\Omega$, which describes the scattering diagram of an individual particle, and the static structure factor $S$, which describes a correlation between particle pairs in the disordered assembly~\cite{vynck2023light}.

Taking pair correlations into account to describe structural correlations in disordered assemblies of identical particles, one can rigorously derive the following expressions for $\ell_\text{s}$ and $g$~\cite{vynck2023light}
\begin{align}\label{eq:ell_s-general}
    \ell_\text{s} = \left[ \rho \int_{4\pi} \frac{\text{d}\sigma_\text{s}}{\text{d}\Omega} (\theta,\varphi) S(\theta) \text{d}\Omega \right]^{-1},
\end{align}
and
\begin{align}\label{eq:g-general}
    g = \frac{\int_{4\pi}  \frac{\text{d}\sigma_\text{s}}{\text{d}\Omega} (\theta,\varphi) S(\theta) \cos \theta \text{d}\Omega}{\int_{4\pi}  \frac{\text{d}\sigma_\text{s}}{\text{d}\Omega} (\theta,\varphi) S(\theta) \text{d}\Omega},
\end{align}
where $\theta$ and $\varphi$ are the scattering angles and $\text{d}\Omega = \sin \theta \text{d}\theta \text{d}\varphi$. These expressions should be evaluated at an effective wavenumber $k_\text{r}=k_0 \text{Re}[n_\text{eff}]$ with $k_0=2\pi/\lambda$ and $n_\text{eff}$ the effective index.

Equations~\eqref{eq:ell_s-general} and \eqref{eq:g-general} were used to obtain the numerical results presented in Figs.~\ref{fig:spectra-exp-vs-num}-\ref{fig:physical-parameters}, where a quantitative comparison with experiments was made. The differential scattering cross-section $\text{d}\sigma_\text{s}/\text{d}\Omega$ was computed here using Mie theory~\cite{bohren2008absorption}. The refractive index of the \ZrO2 particles, $n_\text{p}$, was taken from Wood~\textit{et al.}~\cite{wood1990refractive}, see Fig.~S1 in the Supporting Information, and that of the ambient medium (water) was $n_\text{b}=1.33$. The effective index $n_\text{eff}$ was computed using the coherent potential approximation~\cite{soukoulis1994propagation}. The static structure factor $S$ was evaluated using the Percus-Yevick approximation for hard spheres~\cite{wertheim1963exact}.

Equations~\eqref{eq:ell_s-general} and \eqref{eq:g-general} can be simplified in certain limits. For a dilute particle assembly ($\rho \lambda^3 \ll 1$) with no control over the particle organization, the static structure factor approaches unity ($S=1$) and the effective refractive index $n_\text{eff}$ tends to the refractive index of the background medium $n_\text{b}$. This leads to
\begin{align}\label{eq:ell_s-dilute}
    \ell_\text{s}^{\text{dilute}} = (\rho \sigma_\text{s})^{-1},
\end{align}
and
\begin{align}\label{eq:g-dilute}
    g^{\text{dilute}} = \frac{1}{\sigma_\text{s}} \int_{4\pi} \frac{\text{d}\sigma_\text{s}}{\text{d}\Omega} (\theta,\varphi) \cos \theta \text{d}\Omega,
\end{align}
with the scattering cross-section
\begin{align}\label{eq:sigma-s-general}
    \sigma_\text{s} = \int_{4\pi} \frac{\text{d}\sigma_\text{s}}{\text{d}\Omega} (\theta,\varphi) \text{d}\Omega.
\end{align}
Equations~\eqref{eq:ell_s-dilute} and \eqref{eq:g-dilute} are convenient because, by excluding the impact of correlations that appear at high particle densities, they reveal an invariance property in the optical properties of materials (given the particles) with the reduced particle density $\rho L$. As a matter of fact, high values of $\rho L$ can be reached keeping a low particle density $\rho$, hence with negligible correlations, simply by increasing the sample thickness $L$. These expressions were therefore used to establish the structural color palette in Fig.~\ref{fig:numerical-palette-Rayleigh}(b).

Furthermore, for spherical particles much smaller than the wavelength ($d n_\text{b} / \lambda \ll 1$), which corresponds to the Rayleigh scattering regime of main interest here, the particles behave effectively as radiating electric dipoles, characterized by
\begin{align}\label{eq:diff-sigma-s-ED}
    \frac{\text{d}\sigma_\text{s}^\text{ED}}{\text{d}\Omega}(\theta) = \frac{k_\text{b}^4}{16\pi^2} |\alpha|^2 \frac{1+\cos^2 \theta}{2},
\end{align}
where $k_\text{b} = k_0 n_\text{b}$ is the wavenumber in the background medium and $\alpha$ is the polarizability of the particle. Equation~\eqref{eq:diff-sigma-s-ED} is obtained assuming unpolarized light. Thus, for a dilute, uncorrelated assembly of small particles, we obtain
\begin{align}\label{eq:ell_s-dilute+ED}
    \ell_\text{s}^{\text{dilute}+\text{ED}} = \left[\rho \frac{k_\text{b}^4}{6\pi} |\alpha|^2\right]^{-1},
\end{align}
and
\begin{align}\label{eq:g-dilute+ED}
    g^{\text{dilute}+\text{ED}} = 0.
\end{align}
In the quasi-static limit, the polarizability $\alpha$ of a spherical particle is given by
\begin{align}\label{eq:QS-polarizability}
    \alpha = 4\pi r^3 \frac{\epsilon_\text{p} - \epsilon_\text{b}}{\epsilon_\text{p} + 2 \epsilon_\text{b}},
\end{align}
with the relative permittivities $\epsilon_\text{p}=n_\text{p}^2$ and $\epsilon_\text{b}=n_\text{b}^2$ and $r=d/2$ the particle radius. For non-dispersive materials, the scattering mean free path in Eq.~\eqref{eq:ell_s-dilute+ED} therefore scales exactly as $\lambda^4$. This makes that the structural colors of slabs of Rayleigh-scattering materials now become invariant with the parameter $L/\ell_\text{s}$, when the value of $\ell_s$ is defined at a certain wavelength $\lambda$. A similar reasoning leads to an invariance with the absorption parameter $L/\ell_\text{a}$, supposed constant over the visible range. The numerical results presented in Figs.~\ref{fig:color-palette-absorption} and \ref{fig:carbon-black-solutions} were obtained by varying $L/\ell_\text{s}$ with the value of $\ell_\text{s}$ being evaluated at $\lambda=440~\unit{nm}$ and assuming that it varies exactly as $\lambda^4$, and $L/\ell_\text{a}$ with $\ell_\text{a}$ being constant. These results therefore do not depend specifically on the refractive index and size of the particles, but are strictly valid only in the quasi-static limit and for dilute particle assemblies.

\emph{Monte Carlo light transport simulations:} The basic Monte Carlo method for light transport essentially consists in simulating the trajectories of a large number of random walkers in a finite volume representing the macroscopic material and to estimate the radiometric quantities of interest. Generally assuming that the light transport parameters do not vary with the position in the material, the distribution of step lengths between two scattering events is described by an exponential probability density function $p(l_\text{s})$ with mean $\ell_\text{s}$ and the angular redistribution after a scattering event by the so-called ''phase function''. In all simulations, we used the Henyey-Greenstein phase function, where $g$ is conveniently the only adjustable parameter~\cite{henyey1941diffuse}, to describe the angular redistribution of light upon single scattering events. This is a very reasonable approximation since the $g$ parameter is very close to 0 (i.e., quasi-isotropic scattering) in the Rayleigh scattering regime and the diffuse reflectance and transmittance were integrated angularly. 

The absorption of light is similarly described by an exponential probability density function $p(l_\text{a})$ with mean $\ell_\text{a}$. Random-walk trajectories are then terminated by an absorption process if their length from the entry to the exit is larger than an absorption length $l_\text{a}$. The numerical results presented in Figs.~\ref{fig:color-palette-absorption} and \ref{fig:carbon-black-solutions} were obtained by assuming that the absorption mean free path was constant over the whole visible spectrum.

For each wavelength of interest and each set of parameters, we simulated many random-walk trajectories (typically $10^5$) on slabs of finite thickness $L$ and either infinite or finite lateral dimensions, considering the reflection and refraction due to the refractive index contrast between the matrix and the ambient medium (air, in the present case) via the polarization-averaged Fresnel formulas. The incident light was assumed to be unpolarized and normally incident on the surface. From the random-walk statistics, we computed the spectra for the ballistic transmittance (i.e., the random walkers that would exit the material in transmission without being scattered) and the angularly-integrated diffuse transmittance and reflectance (i.e., the random walkers that would exit the material in transmission or reflection being scattered at least once).

\emph{Conversion between a spectrum and a color:} In all simulations, the spectra were computed for wavelengths between $360~\unit{nm}$ and $830~\unit{nm}$ in steps of $10~\unit{nm}$, which is sufficiently dense to have an accurate prediction of the perceived color. The reflectance and transmittance spectra, normalized to 1, were finally converted to the sRGB color space with a D65 illuminant using a standard procedure~\cite{wyszecki2000color}. To faithfully visualize the predicted colors on a screen, the monitor should thus be calibrated to sRGB.

\subsection*{Materials}

Methyltriethoxysilane (99\%) and vinyltrimethoxysilane (99\%) were obtained from ABCR. Citric acid monohydrate (99\%) was obtained from Merck. Laponite powders was a commercial laponite RD from Laporte Industries. The Yttria-stabilized Zirconia (\ZrO2) particles were obtained in aqueous solutions from Mathym, with a mass concentration of $ 1~\unit{g.cm^{-3}}$ (i.e. 50 wt\% ) and diameter of $20 \pm 3~\unit{nm}$. Figure~S2 in the Supporting Information shows the particle size distribution estimated from the analysis of a TEM image provided by Mathym. Similar colloidal solutions with particles of diameter $90 \pm 15~\unit{nm}$ were used to study the effect of adding black absorbents to the matrix. The carbon graphite particles with an average size of $5~\unit{\um}$ were obtained from Strem Chemicals, see the inset of Fig.~S3 in the Supporting Information for a transmittance electron microscopy image of a particle.

\emph{Doped solution:} The \ZrO2 colloidal solution was first diluted with distilled water, leading to particle densities between $10^2$ and $10^4~\unit{\um^{-3}}$. A powder of carbon graphite particles was used to introduce visible wide band absorption in the liquid samples. Figure~S3 in the Supporting Information shows the ballistic transmittance through an aqueous solution of carbon graphite particles at a concentration of $5 \; 10^{-3}~\unit{g.cm^{-3}}$ in water. The extinction efficiency (accounting for both absorption and scattering) is not perfectly constant in the visible range.

The three aqueous solutions shown in Fig.~\ref{fig:carbon-black-solutions}(b) are: on the left, a solution containing carbon graphite particles only ($5 \; 10^{-3}~\unit{g.cm^{-3}}$ of carbon); on the right, a solution containing 90-\unit{nm}-diameter \ZrO2 particles only ($\rho=4 \; 10^2~\unit{\um^{-3}}$ of particles); in the middle, the same colloidal solution doped with carbon graphite particles ($0.3 \; 10^{-3}~\unit{g.cm^{-3}}$ of carbon). 

\emph{Solid borosilicate clay samples:} The clay was made from a commercially available laponite RD. Laponite is a synthetic trioctahedral smectite of hectorite clay type, whose structure is described in Ref.~\cite{qi1996comportement}. The composition of the laponite is Na$^{0.7+}$[Si$_8$Mg$_{5.5}$Li$_{0.4}$O$_{20}$(OH)$_4$]$^{0.7-}$ and a typical laponite particle generally looks like a disk-shaped platelet (diameter $30~\unit{nm}$, thickness $2~\unit{nm}$). Thanks of its easy high transparency and small crystallite size easily dispersed in water, laponite is particularly suitable for particle doping to produce a structurally-colored material. To prepare the colored clay, the colloidal solution, with a concentration of $5 \; 10^{-4}~\unit{g.cm^{-3}}$ of 20-\unit{nm}-diameter \ZrO2 particles, was first diluted in distilled water heated to $30~\unit{\degreeCelsius}$. The laponite powder was dispersed (3\% solid concentration) in the colloidal solution under strong agitation. The solution was placed in a container and left to rest at room temperature until the gel was obtained. When the solution is dried at higher temperature, gel occurs more quickly but cracks appear. The colored gels can take several months to dry and to be easy to handle.

\textit{Solid hybrid silica-based glass samples:} The hybrid glasses were elaborated by sol-gel process adapted from the protocol described in Refs.~\cite{chateau2012silica, chateau2022plasmonic}. A 1:1 molar mixture of methyltriethoxysilane (MTEOS) and vinyltrimethoxysilane (VTMOS), was hydrolyzed by a solution of citric acid ($15~\unit{g.L^{-1}}$) in water ($n_{\text{H}_2\text{O}}/n_\text{Si}=11$). After hydrolysis, water and alcohol were removed under vacuum. The resulting viscous sol was dissolved in diethyl ether to remove residual water after phase separation. Finally, a solvent exchange was undertaken by addition of ethanol and evaporation of the remaining diethyl ether (final concentration in ethanol : 35\% of dry matter) and filtered on a $0.45~\unit{\um}$ PTFE filter.

The appropriate quantity ($8~\unit{g}$) of the obtained sol was mixed with the required amount of alcoholic colloidal solutions of \ZrO2 particles ($390~\unit{mg}$ of a 40\%wt dispersion in ethanol) in a PFA mold with a screw cap, and the resulting suspension was left to age at room temperature for 2 hours. The suspension was then sonicated in an ultrasonic bath for 10 minutes, and the gelation was initiated by the addition of 3-aminopropyltrimethoxysilane (APTMS, $450~\unit{\uL}$).

The mold was closed with a screw cap and kept undisturbed at room temperature until complete gelation occurs (typically around 20-30 minutes). Finally, the cap of the mold was slightly opened and the material slowly dried in an oven at $45~\unit{\degreeCelsius}$ for one week to yield the final xerogel. The final doped glasses are homogeneous and shaped in solids disks of $3~\unit{cm}$ diameter and $4~\unit{mm}$ thickness. 

\subsection*{Optical measurements}

All optical measurements were done using a UV-Vis-NIR Perkin Elmer (Lambda 1050+) spectrophotometer. Ballistic transmittance was measured in a classical transmission mode, while diffuse transmittance and diffuse reflectance were measured using an integrating sphere adapted to the spectrophotometer. All spectra were recorded in quartz cuvettes in the UV-visible range with a $2~\unit{nm}$ step and an integration time of $0.5~\unit{s}/\text{step}$ taking as a reference quartz cuvette full of water.

\section*{Associated Content}

\subsection*{Data Availability Statement}

Data for this article, including the theoretical structural color palettes and the numerical and experimental spectra, are available on Zenodo.org at DOI: \hyperlink{https://doi.org/10.5281/zenodo.20122650}{10.5281/zenodo.20122650}.

\subsection*{Author contributions}

Project coordination: A.P.; Conceptualization: K.V., A.B.L., C.L.L., S.P., and A.P.; Contextualization in art history: R.T.; Numerical code: K.V.; Numerical simulations: K.V. and C.S.; Material preparation: A.B.L., C.L.L., D.C. and S.P.; Optical measurements: A.B.L., C.L.L., C.S. and A.P.; Data analysis: K.V., A.B.L. C.L.L. and A.P.; Writing -- Original Draft: K.V., A.B.L., C.L.L., S.P. and A.P.; Writing – Review \& Editing: all authors.

\subsection*{Conflicts of interest}

A.P. has a patent on a bitumen painting process for obtaining blue colored layers without blue pigments [FR3078273, filed on the 27th of February 2018]. D.C. is currently employed by Mathym, a company specialized in the development of nanomaterial dispersions.

\section*{Acknowledgements}

A.P. acknowledges Anne Goyer (plastic artist in Avignon, France) for the art-science collaboration around structural blue in paintings.

%\bibliography{references}

%%%%%%% BIBLIOGRAPHY %%%%%%%%%%

\end{document}